\documentclass[12pt,preprint]{aastex}

\usepackage{epsfig}
\usepackage{lscape}
\usepackage{mathptmx}

\newcommand{\dechms}[4]{$#1^{\rm h}#2^{\rm m}#3\mbox{$^{\rm s}\mskip-7.6mu.\,$}#4$}

\newcommand{\Msun}{M$_{\odot}$}

\shorttitle{Molecules in $\eta$ Carinae}
\shortauthors{Loinard et al.}

\begin{document}

\title{Molecules in $\eta$ Carinae}

\author{Laurent Loinard\altaffilmark{1}, Karl M.\ Menten, Rolf G\"usten}

\affil{Max-Planck Institut f\"ur Radioastronomie, Auf dem H\"ugel 69, 53121, Bonn, Germany}

\and

\author{Luis A.\ Zapata, Luis F.\ Rodr\'{\i}guez}

\affil{Centro de Radioastronom\'{\i}a y Astrof\'{\i}sica, Universidad Nacional Aut\'onoma de M\'exico, Apartado Postal 3-72, 58090, Morelia, Michoac\'an, M\'exico}

\altaffiltext{1}{On sabbatical leave from: Centro de Radioastronom\'{\i}a y Astrof\'{\i}sica, Universidad Nacional Aut\'onoma de M\'exico, Apartado Postal 3-72, 58090, Morelia, Michoac\'an, M\'exico}

\begin{abstract}
We report the detection toward $\eta$ Carinae  of six new molecules, CO, CN, HCO$^+$, HCN, HNC, 
and N$_2$H$^+$, and of two of their less abundant isotopic counterparts, $^{13}$CO and H$^{13}$CN. 
The line profiles are moderately broad ($\sim$ 100 km s$^{-1}$) indicating that the emission originates 
in the dense, possibly clumpy, central arcsecond of the Homunculus Nebula. Contrary to previous 
claims, CO and HCO$^+$ do not appear to be under-abundant in $\eta$ Carinae. On the other hand, 
molecules containing nitrogen or the $^{13}$C isotope of carbon are overabundant by about one order 
of magnitude. This demonstrates that, together with the dust responsible for the dimming of $\eta$ Carinae 
following the Great  Eruption, the molecules detected here must have formed {\it in situ} out of CNO-processed
stellar material.
\end{abstract}


\section{Introduction}
$\eta$ Carinae is well known to have experienced a major outburst in the 1840s,
during which it became the second brightest star in the entire sky (e.g., Humphreys 
\& Davidson 1999).  Known as the {\it Great Eruption}, this outburst was associated 
with an episode of extreme mass-loss (about 10 \Msun\ of material was expelled in 
about 20 years) that resulted in the creation of the bipolar-shaped {\it Homunculus 
Nebula} whose current size is about 16$''$ $\times$ 10$''$, or 0.18 $\times$ 0.11 pc
assuming a distance of 2.3 kpc (Walborn 1995, Allen \& Hillier 1993). Over the 
following decades, the visual brightness of $\eta$ Carinae faded by many magnitudes, 
but early infrared observations by Neugebauer \& Westphal (1968) revealed that the 
bolometric luminosity in the second half of the 20$^{th}$ century remained comparable 
to that during the Great Eruption. The dimming at optical wavelengths resulted from 
obscuration by dust particles, presumably formed {\it in situ} out of the ejected material. 

A scant handful of molecular species have been detected toward $\eta$ Carinae. Molecular 
hydrogen, traced by its 2.12 $\mu$m line, appears to be distributed over the outer surface
of the Homunculus Nebula, and is strongest towards the polar caps where the intercepted
column is largest (Smith 2002; 2006). Two other simple diatomic molecules (CH 
and OH) were identified in Hubble Space Telescope STIS spectra through their UV 
absorption lines (Verner et al.\ 2005). Both also originate in the thin outer layer of the 
Homunculus. Finally, radio emission from ammonia (NH$_3$) was detected by Smith 
et al.\ (2006) using the Australia Telescope Compact Array (ATCA). The ammonia emission 
is confined to a region roughly 1 arcsec across, and shares the kinematics of the H$_2$ 2.12 
$\mu$m line in the same region. 

Interestingly, carbon monoxide (CO) has never been detected
toward $\eta$ Carinae in spite of sensitive searches at millimeter, infrared, and UV wavelengths
(Cox \& Bronfman 1995; Smith 2002; Verner et al.\ 2005). As discussed by Smith et al.\ 
(2006), this lack of CO detection could reflect the C/O depletion and N enrichment 
of the material ejected during the Great Eruption. Indeed, the ionized gas surrounding 
the Homunculus nebula is known to be composed of such nitrogen-rich CNO-processed
material (Davidson et al.\ 1982; Davidson et al.\ 1986; Dufour et al.\ 1997; Hillier et al.\ 2001; 
Smith \& Morse 2004). In addition, the 
abundance of the nitrogen-bearing ammonia molecule in the Homunculus itself is 
estimated to be about 2 $\times$ 10$^{-7}$, roughly one order of magnitude higher 
than in cold interstellar clouds (Smith et al.\ 2006). It should be emphasized, however,
that the existing, unsuccessful, searches for CO in $\eta$ Carinae are insufficient to place
meaningful limits on the [CO]/[NH$_3$] abundance ratio in the Homunculus (Smith et al.\
2006). Thus, the low abundance of carbon monoxide in $\eta$ Carinae is still not firmly
established.

The formation and survival of molecules in the harsh environment of $\eta$ Carinae, within
1$''$ (0.01 pc) of a 100 \Msun\ star, remain poorly understood. Other
classes of massive stars (such as red supergiants and Wolf-Rayet stars) are known to 
be surrounded by significant quantities of molecular gas, although at greater distances, 
which might indicate that is represents swept-up ambient interstellar material (Pulliam et 
al.\ 2011; Cappa et al.\ 2001; Rizzo et al.\ 2001, 2003). It is unclear whether
or not there is relation between the mechanisms at work in these objects and those
occurring in the Homunculus. To tackle these issues, it is important to characterize 
the molecular content of the Homunculus, and to determine the physical properties
and spatial distribution of the molecular gas. In this {\it Letter}, we present new sub-millimeter
spectroscopic observations of $\eta$ Carinae designed to search for several new 
molecular species, including carbon monoxide. 

\begin{table}
\caption{Observing log and results}
\centering
\begin{tabular}{lccccccccc}
\hline
Transition & $\nu$ & Rx & $\theta_{mb}$\tablenotemark{a} & $\eta_{mb}$\tablenotemark{b} & W\tablenotemark{c}\\
& (MHz) & & ($''$) & & (K km s$^{-1}$)\\
\hline
CO(3-2)                   & 345795.9899 & {\sc flash}             & 17.5 & 0.73 & 24.0 $\pm$ 3.6 \\%
CO(4-3)                   & 460148.8125 & {\sc flash}             & 13.3 & 0.60 & 31.8 $\pm$ 4.8 \\
CO(6-5)                   & 691473.0763 & {\sc champ$+$}  &  9.0   & 0.48 & 73.5 $\pm$ 11.0 \\%
$^{13}$CO(3-2)     & 330587.9653 & {\sc flash}             & 17.5 & 0.73 & 7.4 $\pm$ 1.1 \\
$^{13}$CO(6-5)     & 661067.2766 & {\sc champ$+$}  & 9.0    & 0.48 & 20.9 $\pm$ 3.1 \\%
CN(3-2)\tablenotemark{d}                & 340247.7700   & {\sc flash}             & 17.5  & 0.73 & 3.7 $\pm$ 0.6\\
HCO$^+$(4-3)       & 356734.2230 & {\sc flash}             & 17.5 & 0.73 & 10.1 $\pm$ 1.5 \\%
HCN(4-3)                & 354505.4773 & {\sc flash}             & 17.5 & 0.73 & 10.7 $\pm$ 1.6 \\%
H$^{13}$CN(4-3)  & 345339.7694 & {\sc flash}             & 17.5 & 0.73 & 8.3 $\pm$ 1.2 \\%
HNC(4-3)                & 362630.3030 & {\sc flash}             & 17.5 & 0.73 & 7.0 $\pm$ 1.0 \\%
N$_2$H$^+$(4-3) & 372672.4645 & {\sc flash}             & 17.5 & 0.73 & 24.5 $\pm$ 3.7 \\%
\hline
\end{tabular}
\tablenotetext{a}{$\theta_{mb}$ is the beam size at each frequency.}
\tablenotetext{b}{$\eta_{mb}$ is the main beam efficiency, used to convert the measured antenna temperatures to main beam temperatures.}
\tablenotetext{c}{The values reported in this column were obtained by integrating over the entire velocity range where emission is detected: W = $\int T_{mb} dv$.}
\tablenotetext{d}{CN(3-2) was not specifically targeted, but happened to be detected near the edge of one of the observed bands. As a consequence, only about half of the hyperfine components were included in the band (see Figure 1), and this observation will have to be repeated in the future.}

\end{table}

\section{Observations}

The observations were performed in 2011 October 12--17 and December 15--20 with the 
Atacama Pathfinder EXperiment telescope (APEX; G\"usten et al.\ 2006) located at an 
altitude of 5100-m on Llano Chajnantor, Chile. The molecular transitions targeted are 
listed in Table 1. Two different receivers were used:  a modified version of the First Light 
Apex Sub-millimeter Heterodyne receiver ({\sc flash}; Heyminck et al.\ 2006) for transitions 
in the 345 and 460 GHz bands, and the Carbon Heterodyne Array of the MPIfR ({\sc champ+}; 
G\"usten et al.\  2008) for transitions in the 690 GHz band. While {\sc flash} is a single-beam
receiver, {\sc champ+} provides spectra simultaneously at 7 positions. Those positions correspond 
to the central (directly on-source) pixel and to 6 lateral points distributed in a hexagonal 
pattern around the central pixel and separated from it by about 19$''$ (G\"usten et al.\ 
2008). The Fast Fourier Transform spectrometer backends provided 32,768 frequency channels, each 76.308 
kHz wide during the {\sc flash} observations, and 8,192 channels, each 183.1 kHz wide 
during the {\sc champ+} observations. This yields velocity resolutions of 0.07--0.08 km 
s$^{-1}$ at all frequencies, but the spectra were Hanning-smoothed to 4--5 km s$^{-1}$ 
during post-processing to improve their signal to noise ratio. 

The observations were obtained in {\sc on-off} position switching mode, with the {\sc off}
position at $\alpha_{J2000.0} = $\dechms{10}{48}{28}{0}, $\delta_{J2000.0}$ = 
\dechms{-59}{25}{45}{0}. This position is known to be devoid of molecular emission. 
Calibration and pointing scans were interspersed with the science spectra throughout the 
observations. The weather conditions and overall system performance were excellent, 
and the resulting spectra of very good quality. As a consequence, very few scans had 
to be discarded, and only low-order polynomial baselines had to be removed. In some 
cases, oscillatory patterns were present in the spectra, and were removed in the
Fourier domain. The intensity scale was converted from T$_A^*$ to T$_{mb}$ using the 
efficiencies listed in Table 1. We estimate the final flux calibration to be accurate to 15\%.

\begin{figure*}[!t]
  \centerline
  {\includegraphics[height=0.85\textwidth,angle=0]{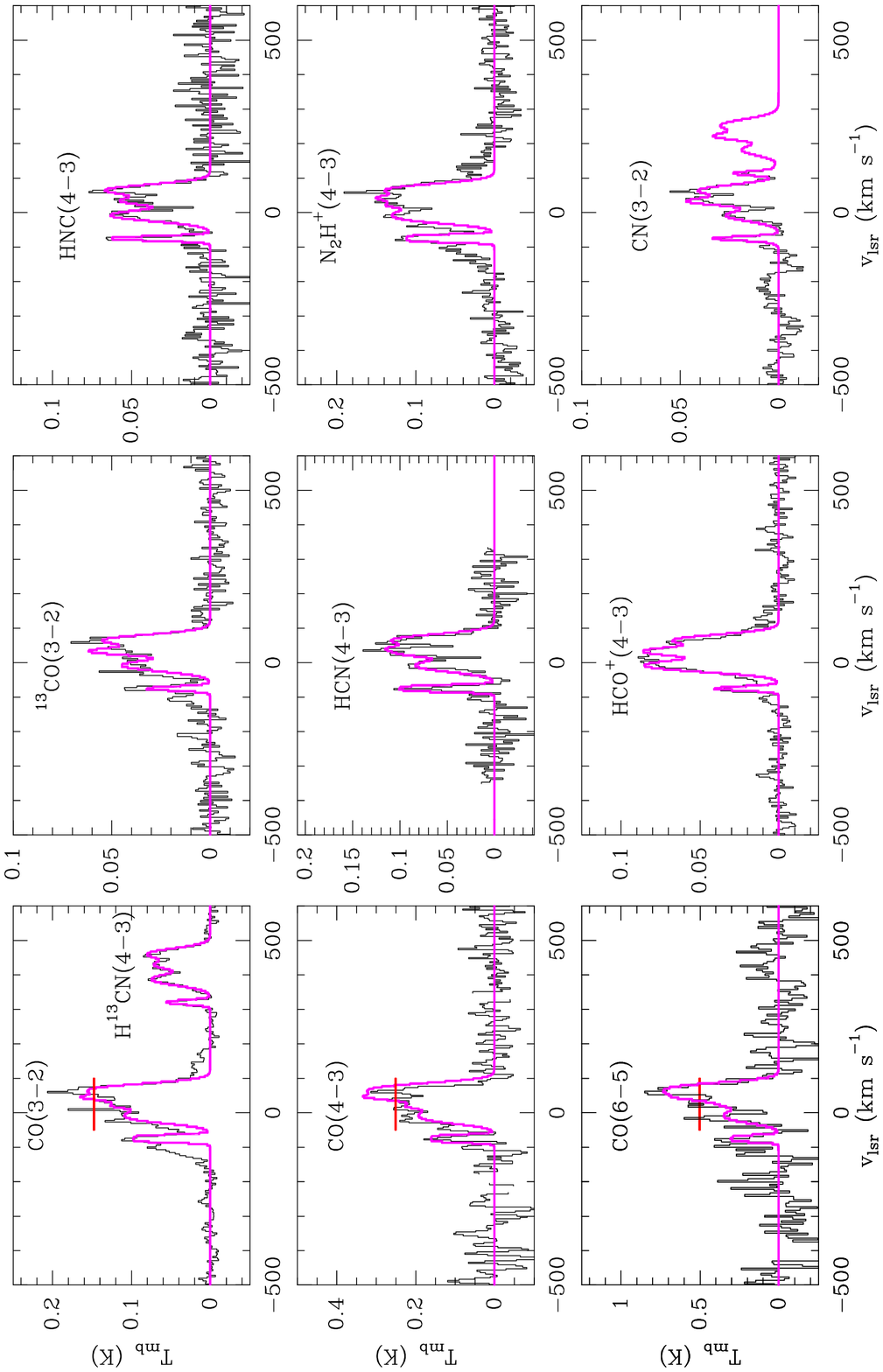}}
  \caption{Observed spectra towards $\eta$ Carinae. The magenta curves show the 
  theoretical spectra expected in the conditions described in the text. The horizontal
  red lines in the three spectra of the left column indicate the typical intensity of the
  CO lines mentioned in Section 3. The CN line was on the edge of the band and
  the corresponding profile misses some hyperfine components.}
\end{figure*}

\section{Results and Analysis}

All the targeted lines were detected toward the source (Figure 1), but no emission was 
seen in the lateral pixels of the {\sc champ+} observations. This demonstrates that the 
molecular emission is confined to the Homunculus itself. In addition, the line profiles 
are quite broad  (up to about $300$ km~s$^{-1}$ full width at zero point,  with ``cores'' of 
roughly $100$ km~s$^{-1}$ full width at half maximum; Figure 1) and reminiscent of the 
NH$_3$ spectra presented by Smith et al.\ (2006). In comparison, the Homunculus has 
expansion velocities of roughly 600 km s$^{-1}$, while the Weigelt knots near the star 
have outward velocities less than 50 km s$^{-1}$ (e.g.\  Hofmann \& Weigelt 1988; Weigelt et 
al.\ 1995; Davidson et al.\ 1995). The similarity between the NH$_3$ spectra (Smith et al.\
2006) and those reported here likely indicates that the emission originates in the central 
few arcseconds of the Homunculus. The emission emission is centered at V$_{lsr}$ 
$\sim$ $+$20 km s$^{-1}$, a value somewhat more positive than the systemic velocity 
of $\eta$ Carinae (--20 km s$^{-1}$;  
Davidson et al.\ 1997; Smith 2004). This is, again, similar to the situation with ammonia 
(Smith et al.\ 2006). The observed profiles are clearly not gaussian. Instead, they exhibit 
significant  sub-structure suggesting that the emission might come from a clumpy material. 
Indeed, all of our spectra are consistent with four velocity components at $v_{lsr}$ = $-$76.2, 
$-$8.9, $+$30.5, and $+$63.5 km s$^{-1}$. Particularly noteworthy is the narrow component 
at $v_{lsr}$ = -- 76.2 km s$^{-1}$ seen in most of our spectra, and most likely associated 
with the strong H$_2$ 1-0 S(1) emission detected by Smith (2004; see also Smith et al.\ 
2006) at the same radial velocity.\footnote{Note that our spectra are measured in the LSR 
rest frame, whereas those in Smith (2002, 2004) and Smith et al.\ (2006) are reported in 
the Heliocentric system. For $\eta$ Carinae, $v_{Hel}$ $\approx$ $v_{lsr}$ $+$ 12 km s$^{-1}$.)}

The combination of observed CO and $^{13}$CO spectra can be used to constrain the
physical conditions of the emitting material. First, we note that the relative peak intensities
of the CO 3$\rightarrow$2, 4$\rightarrow$3, and 6$\rightarrow$5 lines (0.15, 0.25, and 0.5 K; 
see the red marks on Figure 1) are almost exactly in the inverse proportion of the corresponding 
beam areas ($1 \div 1.7 \div 3.8$). This shows that the CO lines are optically thick, and come 
from a region smaller than all the beams (even that at 690 GHz). For such optically thick lines, 
there is a degeneracy between temperature and filling factor. This degeneracy can be removed 
using the $^{13}$CO line intensities, and we find that all the CO and $^{13}$CO lines can be
reproduced for an excitation temperature of order 70 K and a source size of order 1$''$. A higher 
excitation temperature (of, say, 200 K) could reproduce the CO lines provided the source size
were 0.5$''$, but would predict a $^{13}$CO(6-5)/$^{13}$CO(3-2) line ratio of about 8.5, inconsistent
with the observed value of 5. Given the similarities between all the spectra observed here (Figure 
1), it is reasonable to assume that all the molecular emission comes from the same material, so we 
conclude that all the lines reported here originate in a source about 1$''$ in size where the gas is 
at a temperature of order 70 K. 

To estimate the molecular column densities, we used the myXCLASS program\footnote{http://www.astro.uni-koeln.de/projects/schilke/XCLASS} (see Comito
et al.\ 2005 and references therein), and modeled 
the emission as a superposition of the four distinct velocity components identified earlier. We
used line widths for each component consistent with the observed widths of the optically thin
lines ($\Delta v$ = 13.2, 35.8, 21.6, and 36.4 km s$^{-1}$, respectively for the four velocity
components), and found reasonable fits to all the lines for excitation temperatures of 40,
50, 40, and 90 K, respectively for the four components. Our approach to column density
determinations entails a number of approximations. First, the calculations are made under 
the assumption of local thermodynamic equilibrium (LTE). To check that this did not strongly affect our results, we used the publicly 
available non-LTE radiative transfer code {\sc radex} (van der Tak et al.\ 2007) to verify that the 
excitation conditions were consistent with LTE. Secondly, our calculations do not consider 
the opacity due to possible spatial overlap between different velocity components. To decide 
to which extent this problem might affect our conclusions, high angular resolution observations 
will be necessary. The column densities resulting from our analysis are given for each species 
in Table 2, and the corresponding model spectra are shown in Figure 1. We note that 
the CO column density itself is somewhat uncertain due to its substantial opacity. 

The abundance of each species was calculated relative to CO and to molecular hydrogen, assuming 
a column density N(H$_2$) = 3 $\times$ 10$^{22}$ cm$^{-2}$ as estimated to be appropriate 
for this part of the Homunculus by Smith et al.\ (2006). We note, however, that a somewhat larger 
column density of H$_2$ might also be plausible. Based on sub-millimeter and far-infrared observations, 
Gomez et al.\ (2010) recently determined the mass of dust surrounding $\eta$ Carinae to 
be about 0.4 \Msun. For a standard gas-to-dust ratio of 100, this would yield an {\em average}
H$_2$ column density of about 2 $\times$ 10$^{23}$ cm$^{-2}$. If the dust distribution were clumpy, 
however, the column density appropriate for the individual clumps might be several times larger,
and the abundances quoted in Table 2 could be proportionately lower.

\section{Discussion}

Within its uncertainty, the abundance of CO derived here for $\eta$ Carinae is similar to its canonical 
interstellar value of 10$^{-4}$. It is also similar to the typical CO abundances found in O-rich circumstellar
envelopes (Ziurys et al.\ 2009). Thus, CO does not appear to be under-abundant in the Homunculus
Nebula, contrary to previous claims based on unsuccessful CO searches. The abundance of HCO$^+$ 
is similar to its value in dense massive cores ($\sim$ 2 $\times $10$^{-8}$; Vasyunina et al.\ 2011) and 
in the dense envelopes surrounding low-mass stars ($\sim$ 1.2 $\times$ 10$^{-8}$; Hogerheijde et al.\ 
1997). It is also in the mid-range of observed abundances in evolved stars with oxygen-rich circumstellar 
envelopes (0.5--13 $\times$ 10$^{-8}$; Pulliam et al.\ 2011), but significantly larger than the abundance
in carbon-rich stars, such as IRC$+$10216 (where it is 4$\times$ 10$^{-9}$; Pulliam et al.\ 2011). 
We conclude that both CO and HCO$^{+}$ have roughly standard abundances in the Homunculus. 

The nitrogen-bearing molecules, on the other hand, are found to be highly over-abundant in $\eta$
Carinae. While the average abundance of HCN and HNC in low- and high-mass dense cores is 2--7 
$\times$ 10$^{-9}$ (Vasyunina et al.\ 2011), their abundances in $\eta$ Carinae are 0.7--2 $\times$ 
10$^{-7}$ (Table 2). Similarly, the abundance of N$_2$H$^+$ is about two orders of magnitude higher 
in the Homunculus than in dense cores (where it is, on average, 2 $\times$ 10$^{-9}$; Vasyunina et al.\ 
2011). The situation is much the same for ammonia (Smith et al.\ 2006) showing that nitrogen-bearing 
molecules are consistently one order of magnitude more abundant in the Homunculus than in the dense 
interstellar medium. Although chemical effects might affect the abundance of specific individual molecules, 
this combination of results suggests that the abundance of nitrogen itself must be  enhanced by one order 
of magnitude in the Homunculus. The comparison between the abundances of N-bearing species in
$\eta$ Carinae and those in O-rich circumstellar envelopes is somewhat confusing. While HCN is
about one order of magnitude less abundant in $\eta$ Carinae than in O-rich envelopes, the abundance
of HNC is similar in both kinds of objects. As a consequence, the [HCN]/[HNC] ratio in $\eta$ Carinae is 
of the order of a few, similar to its value, of order unity, in quiescent interstellar cores (Padovani et al.\ 2011), 
but very different from its value (a few hundred) in oxygen-rich envelopes (Ziurys et al.\ 2009). CN, on the
other hand, is about one order of magnitude more abundant in $\eta$ Carinae than in O-rich envelopes
(Ziurys et al.\ 2009).

An important conclusion of our observations concerns the relative abundance of isotopic
forms of carbon. The [HCN]/[H$^{13}$CN] ratio is estimated to be about 2, while the [CO]/[$^{13}$CO] ratio
is of order 5. This is much smaller than the interstellar $^{12}$C/$^{13}$C isotopic ratio at the galactocentric 
radius of $\eta$ Carinae ($\sim$ 70; Milam et al.\ 2005). On the other hand, such a low value of the 
$^{12}$C/$^{13}$C is an expected result of the CNO cycle. In particular, for the CNO cycle at a temperature of
10$^8$ K, the expected equilibrium value of the $^{12}$C/$^{13}$C ratio is of order 4 (Rose 
1998, Chap. 6), in very good agreement with the isotopic ratio measured here. The high abundance of 
nitrogen in the ionized gas surrounding the Homunculus (Smith \& Morse 2004) and in the Homunculus 
itself (as documented above) are also expected consequences of the CNO process. Thus, the molecular 
observations presented here strongly support the idea that the material expelled during the Great Eruption 
is CNO-processed stellar matter. 

\begin{table}[!tb]
\caption{Estimated column densities and abundances}
\centering
\begin{tabular}{lrrr}
\hline
Species & N (cm$^{-2}$) & N/N(H$_2$)\tablenotemark{a} & N/N(CO)\\
\hline
CO & 6.5 $\times$ 10$^{18}$ & 2.2 $\times$ 10$^{-4}$ & 1 \\
$^{13}$CO & 1.4 $\times$ 10$^{18}$ & 4.7 $\times$ 10$^{-5}$ & 2.2 $\times$ 10$^{-1}$\\
CN & 9.0 $\times$ 10$^{15}$ & 3.0 $\times$ 10$^{-7}$ & 1.4 $\times$ 10$^{-3}$\\
HCO$^+$ & 1.7 $\times$ 10$^{15}$ & 5.7 $\times$ 10$^{-8}$ & 2.6 $\times$ 10$^{-4}$\\
HCN & 5.5 $\times$ 10$^{15}$ & 1.8 $\times$ 10$^{-7}$ & 8.5 $\times$ 10$^{-4}$\\
H$^{13}$CN & 3.1 $\times$ 10$^{15}$ & 1.0 $\times$ 10$^{-7}$ & 4.8 $\times$ 10$^{-4}$\\
HNC & 2.1 $\times$ 10$^{15}$ & 7.0 $\times$ 10$^{-8}$ & 3.2 $\times$ 10$^{-4}$\\
N$_2$H$^+$ & 6.1 $\times$ 10$^{15}$ & 2.0 $\times$ 10$^{-7}$ & 9.4 $\times$ 10$^{-4}$\\
\hline
\end{tabular}
\tablenotetext{a}{Abundance of each species relative to H$_2$, assuming N(H$_2$) = 3 $\times$ 10$^{22}$ cm$^{-2}$ (Smith et al.\ (2006).}
\end{table}

We mentioned in the Introduction that dust grains must have formed out of the material ejected by $\eta$
Carinae during the Great Eruption. The present results demonstrate that large quantities of molecular 
material have also formed out of this material. It will be interesting to analyze the chemistry that led to 
the formation of these molecules from the theoretical standpoint, because the elemental composition of the
gas (particularly the N and $^{13}$C enrichment) and the physical conditions (especially the strong
UV field) are very different from those in the interstellar gas. Additionally, the chemistry at play occurred 
in just a few decades. From the observational point of view, it will
be important to further characterize the molecular content of $\eta$ Carinae. Searching for additional
nitrogen-bearing molecules such as HC$_3$N would be particularly interesting. To further characterize
the isotopic composition of the gas, it would also be important to search for molecules containing the $^{15}$N,
$^{17}$O, and $^{18}$O isotopes because CNO nucleosynthesis models make specific predictions for the relative
abundance of these elements. Finally, it would be very interesting to characterize the spatial distribution of the
molecular material in the Homunculus. Our observations suggest a source size of order 1$''$, but it is clear from
the composite nature of the line profiles, that observations at sub-arcsecond resolution would enable a 
detailed study of the spatial distribution of the molecular material and of its kinematics. ALMA will, of course,
be the instrument of choice for such observations.

\section{Conclusions and perspectives}

In this {\it Letter}, we have reported the detection of six new molecules, including carbon
monoxide, and two of their less abundant isotopic forms toward $\eta$ Carinae. This triplicates 
the number of molecules known in this object. While the abundances of CO and HCO$^+$ are 
found to be standard, molecules containing nitrogen or the $^{13}$C isotopic form of carbon are 
over-abundant by about one order of magnitude. This indicates that the material expelled by $\eta$ 
Carinae during the Great Eruption is CNO-processed stellar matter.

Additional single-dish and interferometric observations will be very important to further
characterize the chemical composition of the gas on the Homunculus, and to establish
its spatial distribution. Observations of additional nitrogen bearing molecules and of
species containing specific isotopes of carbon, oxygen, and nitrogen will be 
particularly interesting. Herschel spectroscopic observations, currently being collected,
will also provide very interesting, complementary information.

\acknowledgments We thank Arnaud Belloche, Antoine Gusdorf, and Friedrich Wyrowski 
for their help with the observations and data reduction. L.L., L.A.Z., and L.F.R.\ acknowledge the 
support of DGAPA, UNAM, and of CONACyT (M\'exico). LL is indebted to the Alexander 
von Humboldt Stiftung for financial support. This research made use of the myXCLASS 
program (https://www.astro.uni-koeln.de/projects/schilke/XCLASS), which accesses 
the CDMS (http://www.cdms.de) and JPL (http://spec.jpl.nasa.gov) molecular data
bases.


\begin{thebibliography}{}

\bibitem[Allen 
\& Hillier(1993)]{1993PASAu..10..338A} Allen, D.~A., \& Hillier, D.~J.\ 1993, Proceedings of the Astronomical Society of Australia, 10, 338 

\bibitem[Cappa et al.(2001)]{2001AJ....121.2664C} Cappa, C.~E., Rubio, M., 
\& Goss, W.~M.\ 2001, \aj, 121, 2664 

\bibitem[Comito et al.(2005)]{2005ApJS..156..127C} Comito, C., Schilke, P., 
Phillips, T.~G., et al.\ 2005, \apjs, 156, 127 

\bibitem[Cox 
\& Bronfman(1995)]{1995A&A...299..583C} Cox, P., \& Bronfman, L.\ 1995, \aap, 299, 583 

\bibitem[Davidson et al.(1986)]{1986ApJ...305..867D} Davidson, K., Dufour, 
R.~J., Walborn, N.~R., \& Gull, T.~R.\ 1986, \apj, 305, 867 

\bibitem[Davidson et al.(1997)]{1997AJ....113..335D} Davidson, K., Ebbets, 
D., Johansson, S., et al.\ 1997, \aj, 113, 335 

\bibitem[Davidson et al.(1982)]{1982ApJ...254L..47D} Davidson, K., Walborn, 
N.~R., \& Gull, T.~R.\ 1982, \apjl, 254, L47

\bibitem[Dufour et al.(1997)]{1997ASPC..120..255D} Dufour, R.~J., Glover, 
T.~W., Hester, J.~J., et al.\ 1997, Luminous Blue Variables: Massive Stars 
in Transition, 120, 255 

\bibitem[Gomez et al.(2010)]{2010MNRAS.401L..48G} Gomez, H.~L., Vlahakis, 
C., Stretch, C.~M., et al.\ 2010, \mnras, 401, L48 

\bibitem[G{\"u}sten et 
al.(2006)]{2006A&A...454L..13G} G{\"u}sten, R., Nyman, L.~{\AA}., Schilke, P., et al.\ 2006, \aap, 454, L13 

\bibitem[G{\"u}sten et al.(2008)]{2008SPIE.7020E..25G} G{\"u}sten, R., 
Baryshev, A., Bell, A., et al.\ 2008, \procspie, 7020,  

\bibitem[Heyminck et 
al.(2006)]{2006A&A...454L..21H} Heyminck, S., Kasemann, C., G{\"u}sten, R., de Lange, G., \& Graf, U.~U.\ 2006, \aap, 454, L21 

\bibitem[Hofmann 
\& Weigelt(1988)]{1988A&A...203L..21H} Hofmann, K.-H., \& Weigelt, G.\ 1988, \aap, 203, L21

\bibitem[Hogerheijde et al.(1997)]{1997ApJ...489..293H} Hogerheijde, M.~R., 
van Dishoeck, E.~F., Blake, G.~A., 
\& van Langevelde, H.~J.\ 1997, \apj, 489, 293 

\bibitem[Hillier et al.(2001)]{2001ApJ...553..837H} Hillier, D.~J., 
Davidson, K., Ishibashi, K., \& Gull, T.\ 2001, \apj, 553, 837 

\bibitem[Humphreys 
\& Davidson(1999)]{1999ASPC..179....3H} Humphreys, R.~M., \& Davidson, K.\ 1999, Eta Carinae at The Millennium, 179, 3 

\bibitem[Milam et al.(2005)]{2005ApJ...634.1126M} Milam, S.~N., Savage, C., 
Brewster, M.~A., Ziurys, L.~M., \& Wyckoff, S.\ 2005, \apj, 634, 1126 

\bibitem[Neugebauer 
\& Westphal(1968)]{1968ApJ...152L..89N} Neugebauer, G., \& Westphal, J.~A.\ 1968, \apjl, 152, L89 

\bibitem[Padovani et 
al.(2011)]{2011A&A...534A..77P} Padovani, M., Walmsley, C.~M., Tafalla, M., Hily-Blant, P., \& Pineau Des For{\^e}ts, G.\ 2011, \aap, 534, A77 

\bibitem[Pulliam et al.(2011)]{2011ApJ...743...36P} Pulliam, R.~L., 
Edwards, J.~L., \& Ziurys, L.~M.\ 2011, \apj, 743, 36 

\bibitem[Rizzo et al.(2003)]{2003IAUS..212..740R} Rizzo, J.~R., 
Mart{\'{\i}}n-Pintado, J., 
\& Desmurs, J.-F.\ 2003, A Massive Star Odyssey: From Main Sequence to Supernova, 212, 740 

\bibitem[Rizzo et al.(2001)]{2001ApJ...553L.181R} Rizzo, J.~R., 
Mart{\'{\i}}n-Pintado, J., \& Henkel, C.\ 2001, \apjl, 553, L181 

\bibitem[Rose(1998)]{1998asa..book.....R} Rose, W.~K.\ 1998, Advanced 
Stellar Astrophysics, by William K.~Rose, pp.~494.~ISBN 
0521581885.~Cambridge, UK: Cambridge University Press, May 1998.

\bibitem[Smith(2004)]{2004MNRAS.351L..15S} Smith, N.\ 2004, \mnras, 351, 
L15 

\bibitem[Smith 
\& Morse(2004)]{2004ApJ...605..854S} Smith, N., \& Morse, J.~A.\ 2004, \apj, 605, 854 

\bibitem[Smith et al.(2004)]{2004ApJ...605..405S} Smith, N., Morse, J.~A., 
Gull, T.~R., et al.\ 2004, \apj, 605, 405 

\bibitem[Smith(2002)]{2002MNRAS.337.1252S} Smith, N.\ 2002, \mnras, 337, 
1252 

\bibitem[Smith et al.(2006)]{2006ApJ...645L..41S} Smith, N., Brooks, K.~J., 
Koribalski, B.~S., \& Bally, J.\ 2006, \apjl, 645, L41 

\bibitem[Smith(2006)]{2006ApJ...644.1151S} Smith, N.\ 2006, \apj, 644, 1151 

\bibitem[van der Tak et 
al.(2007)]{2007A&A...468..627V} van der Tak, F.~F.~S., Black, J.~H., Sch{\"o}ier, F.~L., Jansen, D.~J., \& van Dishoeck, E.~F.\ 2007, \aap, 468, 627 

\bibitem[Vasyunina et 
al.(2011)]{2011A&A...527A..88V} Vasyunina, T., Linz, H., Henning, T., et al.\ 2011, \aap, 527, A88 

\bibitem[Verner et al.(2005)]{2005ApJ...629.1034V} Verner, E., Bruhweiler, 
F., Nielsen, K.~E., et al.\ 2005, \apj, 629, 1034 

\bibitem[Walborn(1995)]{1995RMxAC...2...51W} Walborn, N.~R.\ 1995, Revista 
Mexicana de Astronomia y Astrofisica Conference Series, 2, 51 

\bibitem[Weigelt et al.(1995)]{1995RMxAC...2...11W} Weigelt, G., Albrecht, 
R., Barbieri, C., et al.\ 1995, Revista Mexicana de Astronomia y 
Astrofisica Conference Series, 2, 11 

\bibitem[Ziurys et al.(2009)]{2009ApJ...695.1604Z} Ziurys, L.~M., 
Tenenbaum, E.~D., Pulliam, R.~L., Woolf, N.~J., 
\& Milam, S.~N.\ 2009, \apj, 695, 1604 


\end{thebibliography}
\end{document}